\documentclass{elsart}
\usepackage{graphicx,hhline,tabularx}
\usepackage{amssymb}
\usepackage{amsmath}
\begin{document}

\begin{frontmatter}
\title{Transient Information Flow in a Network of Excitatory and Inhibitory Model Neurons: Role of Noise and Signal
Autocorrelation}
\author{Julien Mayor\corauthref{cor1}}
\author{}
\ead{julien.mayor@epfl.ch}
\ead[url]{http://diwww.epfl.ch/$\sim$jmayor}
\corauth[cor1]{Corresponding author}
\author{and Wulfram Gerstner}
\address{Laboratory of Computational Neuroscience LCN \\
Brain and Mind Institute \\
Swiss Federal Institute of Technology Lausanne\\
CH-1005 Lausanne EPFL}
\end{frontmatter}
\pagebreak

\section*{Abstract}
We investigate the performance of sparsely-connected networks of integrate-and-fire neurons for ultra-short term
information processing.
We exploit the fact that the population activity of networks with balanced excitation and inhibition can switch from
an oscillatory firing regime to a state of asynchronous irregular firing or quiescence depending on the rate
of external background spikes.

We find that in terms of information buffering the network performs best for a moderate, non-zero, amount of noise.
Analogous to the phenomenon of stochastic resonance the performance decreases for higher and lower noise levels.
The optimal amount of noise corresponds to the transition zone between a quiescent state and a regime of
stochastic dynamics.
This provides a potential explanation on the role of non-oscillatory population activity in a simplified model of
cortical micro-circuits.
\vfill
\pagebreak

\section{Introduction}
The brain processes information in a constantly varying world. 
The visual scenes one sees in everydaylife are extremely rich 
and change rapidly in time. In addition, the human eye performs more 
than three saccades every second causing sudden changes in the visual input \cite{ORegan92}. 
Complex sounds, such as speech or music vary continuously 
in time and frequency. Moreover, the human brain has to deal 
with simultaneous sound sources from different origins, that are superimposed and mixed together.
From such a perspective, it is clear that
cortical areas are confronted with time-varying sensory inputs rather
than stationary stimuli. 
While attractor neural networks
are considered suitable as models of working memory \cite{Wang01,Fusi02a,Amit97a},
they are less useful to explain ultra-short memory buffer properties
in the range of 10 to 100ms of signals that vary continuously in space or time.
Recently, models of continuous information processing in
recurrent neural networks have been proposed under the names of
Liquid State Machine (LSM, \cite{MaassETAL:01a}) and echo state
networks \cite{Jaeger02,Jaeger04}, both related to the timing network of Buonomano and Merzenich \cite{Buonomano95}.
Those models, that perform computation using transient activity (as opposed to convergence to a stable state), 
are sometimes referred to as models of computation with dynamic states.
The idea underlying those models is
that the instantaneous state of the network provides a rich
reservoir of non-linear spatio-temporal transformations of the
inputs. 

In the framework of the liquid state machines, learning only
acts on the readout structures, the network itself remaining
fixed. Several recent
studies focus either on the
biological realism of such model networks
or on possible implementations of this approach in machine
learning or robotics. 
The role played by dynamical synapses was studied from the perspective of efficient temporal processing in
\cite{NatschlaegerETAL:01,NatschlaegerETAL:01a}.
Principles of liquid state machines were applied to the analysis of a variety of time series 
\cite{NatschlaegerETAL:02,NatschlaegerETAL:03} for
computer vision \cite{MaassETAL:02a}, movement generation and control \cite{JoshiMaass:03,LegensteinETAL:03},
and prediction of chaotic time series \cite{Jaeger04}.
The idea of performing (simple) computations based on perturbation of a real liquid (water) was investigated in
\cite{Fernando03}.

Cortical micro-circuits are extremely complex recurrent networks of neurons.
A given neuron is functionally connected to only a relatively small fraction of the other neurons.
The connection strength between neurons is not fixed, but is mediated by
synapses that have their own dynamics \cite{Braitenberg91,Cajal09,Douglas98,Defelipe93,Gupta00,Silberberg02}.
Maass and colleagues have chosen to simulate networks with a detailed set of
biologically-inspired parameters. They implemented distance-dependent connectivity and 
different refractory periods and
thresholds for the inhibitory and excitatory pools. They also
investigated the role of dynamic synapses and introduced
stochasticity in the values of some parameters
\cite{MaassETAL:01a}.

On the other end of the spectrum of neural network models are idealised
networks that replace real cortical connectivity by sparse random connections
and reduce neuronal diversity to two neuron types, i.e. excitatory and inhibitory.
Even from these simplified models, a rich dynamics can emerge
such as fast and slow oscillations, synchrony
and even chaotic behaviour \cite{Brunel00,Brunel99,Hansel96,Sompolinsky88,Vreeswijk96}.
These simplified models have the advantage that theoretical methods for investigating the dynamics are available
via an analytical approach, so that the parameter space can be explored systematically. 
Based on methematical analysis, the network behaviour can be
classified in a small number of types, such as fast oscillations, slow oscillations, or spontaneous asynchronous
activity \cite{Brunel00,Ermentrout98}, a rigorous classification which would be difficult to obtain by pure simulation-based studies.

A remarkable result of such studies concerns asynchronous activity \cite{Brunel00,Brunel99}: in the phase diagram
of the activity of sparsely-connected network of inhibitory and excitatory neurons, there exists a stable phase of asynchronous
irregular firing in which the overall activity is stationary (its statistical properties such as mean and 
variance are time-independent) but the activity of individual neurons is highly irregular.
From the point of view of information processing in a cortical-like network, 
such a regime of firing provides very interesting properties.
In particular, we can imagine that information about past input can be buffered
in the perturbations of the stable state of asynchronous activity.

The presence of such a memory trace of past stimulations is hypothesised for example in 
models of conditioning (such as reinforcement learning \cite{Sutton98}) under the name of eligibility trace. 
However the underlying mechanism for such a trace is not known, even though many models have blossomed in the past 
decade. 
In most models (for example \cite{Suri98}), the clock is replaced by a series of spectra without any strong 
biological relevance. Bullock and al. \cite{Brown99} have a model of a clock that is 
dependent on the $Ca^{2+}$ activity and Contreras-Vidal \cite{Contreras99} hypothesised an activation 
of different subsets of striosomes. However these mechanisms are supposed to cover 
many orders of magnitude in timing (from a fraction of a second up to a few seconds), 
and it is unlikely that such a broad spectrum can be covered by a single ionic mechanism.
Another way of keeping track of time is to have a recurrent neural network,
that will store dynamically the timing information. Such a mechanism is likely to exist in the olivo-cerebellar system
\cite{Kistler02}.

In this paper we make use of the mathematical analysis of sparsely-connected networks \cite{Brunel00}
in order to relate macroscopic states of the network with evaluations of information buffering, 
thus establishing a link between network theory and the liquid state machines \cite{MaassETAL:01a}.
Specifically, we inject a time-dependent input current into a network of
excitatory and inhibitory neurons with sparse connectivity.
We want to know for how long information about an input signal is kept in the network.
In order to get a quantitative answer, we attempt to reconstruct the input signal at time $t-T$
using the instantaneous membrane potentials of all neurons at time $t$.
If the error of signal reconstruction is small, we say that the network is capable of
"buffering" information for a time $T$.

\section{Model and Methods}
\subsection{Integrate-and-fire neuron}
The system we study is a sparsely connected network of leaky
integrate-and-fire (IF) neurons. Our network is composed of $n=200$
IF neurons, 80\% of which are excitatory and 20\% inhibitory. The
equation for the membrane potential of such neurons can be
written as:
\begin{equation}\label{if}
    \tau_m \dot{u_i}(t) = -u_i(t) + RI_i(t)
\end{equation}
where $I_i$  is the total input, $\tau_m = RC_m$ the effective
membrane time constant, $R$ the effective input resistance and
$C_m$ the membrane capacity. The total input of neuron $i$ can be
decomposed into contributions of the presynaptic spikes from
other neurons $j$ within the micro-circuit under consideration, an
external drive $I^{ext}$ and a noise term $I^{noise}$ that models background
spikes (described in detail in section \ref{stochsection}) from other brain areas that are not described explicitly:
\begin{equation}\label{synint}
    RI_i(t)=\tau_m \sum_{j \in M_i} \omega_j \sum_k \delta(t-t_j^k-D) + RI_i^{ext}(t)+ RI_i^{noise}(t) 
\end{equation}
where $M_i$ is the ensemble of presynaptic neurons,
$t_j^k$ the time neuron $j$ fires its $k$'th spike
and $D$ is a transmission delay.
Synaptic weights are chosen from two values, either excitatory,
$\omega_j=\omega_E$ or inhibitory, $\omega_j=-\omega_I$.
Thus from equations (\ref{if}) and (\ref{synint}), the EPSP's resulting from
spike arrivals at an excitatory synapse are of amplitude $\omega_E$
and decay with the membrane time constant $\tau_m$. Analogously, the
IPSP's resulting from
spike arrivals at an inhibitory synapse are of amplitude $-\omega_I$
and decay with the membrane time constant $\tau_m$.
The total depolarisation of neuron $i$ is the sum of all EPSP's and IPSP's
that have arrived since the last spike of postsynaptic neuron $i$.

When the depolarisation $u_i(t)$ of neuron $i$ reaches a threshold $\theta$,
a spike is emitted
and the membrane potential is reset to a potential
$u_{reset}$ after an absolute refractory period $\tau_{rp}$ during
which the neuron is insensitive to any stimulation. The numerical values are shown
in table 1.

\begin{table}
\hbox{
\begin{tabularx}{60mm}{|X|c|c|}
\multicolumn{3}{l}{\bf{Neuron parameters}} \\ \hhline{===}
Membrane time constant & $\tau_m$ & 20 ms \\ \hline Threshold &
$\theta$ & 5 mV \\ \hline Absolute refractory period &
$\tau_{rp}$ & 2 ms \\ \hline Transmission delay & $D$ & 1 ms \\
\hline Reset membrane potential & $u_{reset}$ & 0 mV \\ \hline
Effective input resistance & $R$ & 10 M$\Omega$ \\ \hline
Membrane capacity & $C_m$ & 2 nF \\ \hline
\end{tabularx}
\hspace{3mm}
\begin{tabularx}{71mm}{|X|c|c|}
\multicolumn{3}{l}{\bf{Network parameters}} \\ \hhline{===}
Number of neurons & $n$ & 200 \\ \hline Connectivity ratio &
$\epsilon$ & 0.2 \\ \hline Excitatory synaptic efficacy &
$\omega_E$ & 1 mV \\ \hline Inhibitory synaptic efficacy &
$\omega_I$ & 5 mV \\ \hline Number of external excitatory neurons
& $N_E$ & 100 \\ \hline Spiking freq. of the external excitatory
neurons (model A) & $\nu_{exc}$ & $10^{-5}$-100 Hz \\ \hline
\end{tabularx}
\vspace{1cm}}
\caption{Left. Numerical values of the model neurons (leaky Integrate-and-Fire neuron) used for the simulations. 
Voltage is measured with respect to the resting potential. 
Right. Numerical parameters of the simulated neural network. The connection probability of 
one neuron to another corresponds to the connectivity ratio $\epsilon$.}
\end{table}

\subsection{Stochastic background input}\label{stochsection}
We consider two noise models. The first model assumes stochastic 
arrival of \emph{excitatory} spikes from other areas. Since a complete
analytical description of the impact of this noise model on the
dynamics of the network has been done \cite{Brunel00}, we can connect
our results to the known macroscopic dynamics of the randomly connected networks.

In a second noise model, we include 
both excitatory \emph{and} inhibitory spikes from outer regions.
The combination of inhibition with excitation generalises
the first model and permits change in the fluctuations of the drive without changing its mean.
It also allows us to
draw an interesting link to a physical phenomena known under the name of \emph{stochastic resonance}.

\subsubsection{Model A: Stochastic spike arrival at excitatory synapses}
In order to simulate the intense synaptic bombardment of neurons observed
\emph{in vivo} (for a review see \cite{Destexhe03}),
we first consider purely excitatory stochastic
spike arrivals coming from $N_E$ external excitatory neurons.
We assume that each spike changes instantaneously the membrane
potential of the post-synaptic neurons (satisfying equation (\ref{if})).
Stochastic spike arrivals can be described by the sum of $N_E$ homogeneous
Poisson processes of rate $\nu_{exc}$. In such a description the
probability of having a spike in a time window is stationary in
time. There is no correlation between spikes, i.e. the occurrence
of a spike at a given time does not influence the future. If we
make the approximation that a neuron receives a large number of
presynaptic contributions per unit time, each generating a change
$\omega_E$ in the membrane potential that is relatively
small compared to the firing threshold $\theta$,
the noise term can be written as a constant drive plus a
fluctuating part\footnote{This substitution is known as a diffusion approximation, see e.g. \cite{Brunel00}}:
\begin{equation}
    RI_i^{noise}(t) =  \mu_E +  \sigma_E \sqrt{\tau_{m}} \eta_i(t)
\end{equation}
where $\mu_E=N_E \omega_E \nu_{exc} \tau_m$ and $\sigma_E=\omega_E \sqrt{N_E \nu_{exc}\tau_m}$.
By $\eta$, we denote standard Gaussian white noise of zero mean and unit variance, i.e.,
\begin{gather}
    <\eta_i(t)>=0   \notag \\
    <\eta_i(t)\eta_i(t')>=\delta(t-t') \notag
\end{gather}
This type of excitation correspond to the analysis of
asynchronous activity in sparsely connected network of IF neurons
made in \cite{Brunel00}. The results from our simulations can
thus be directly interpreted with the phase diagram of the
activity, figure 2 in reference \cite{Brunel00}.

We will focus on the influence of the external spiking activity
to the performance of the network in the range
$10^{-5}Hz<\nu_{exc}<100Hz$.

\subsubsection{Model B: Excitatory and inhibitory stochastic spike arrivals}\label{model2}
In the second noise model we assume stochastic spike arrival from
both excitatory and inhibitory neurons. Let us consider that we
have an external population $N_E$ of excitatory neurons firing at
a rate $\nu_{exc}$, as in model $A$. We now add a further input population 
(referred thereafter as the "balanced population") made out of two groups
of neurons, one of them excitatory and the other one inhibitory.
In order to have a balance between excitation and inhibition, we
choose to have five times as many excitatory neurons as
inhibitory neurons, but with inhibitory synapses that are five
times as strong, i.e.,
$N_E^{pop}=\frac{\omega_I}{\omega_E}N_I^{pop}$, with
$\frac{\omega_I}{\omega_E}=5$. This approximate balance between
excitation and inhibition is thought to take place at a
functional level in cortical areas \cite{Liu04,Shu03}. 
The choice we made to have an exact balance in this
additional population of neurons is for the sake of simplicity
only. Its mean contribution is therefore zero and a change in the
firing rate of this population only affects the variance of the
drive. In addition, modulation of the variance part of the drive will allow 
us to focus on pure noise effects for a fixed mean drive.

With such a model we can write the noise term, similar to the
previous model A:
\begin{equation}\label{baleq}
    RI_i^{noise}(t) = \mu_E + \sigma_{pop} \sqrt{\tau_{m}}\eta_i(t)
\end{equation}
where we have defined $\mu_E=N_E \omega_E \nu_{exc} \tau_m$ and \\
$\sigma_{pop}=\sqrt{\tau_m [N_E \omega_E^2 \nu_{exc} + \nu_{pop}(N_E^{pop}\omega_E^2 + N_I^{pop}\omega_I^2)]}$.

The effective mean drive corresponds to the mean excitation from
the purely excitatory population only since the mean contribution
of the balanced population is zero . The variance can now be
varied independently of the effective mean input, by changing the
discharge rate $\nu_{pop}$ of the balanced population.

The noise terms in both models (A) and (B) can be expressed as:
\begin{equation}
    RI_i^{noise}(t) = \mu_E + \sigma \sqrt{\tau_{m}} \eta_i(t) \notag
\end{equation}
where $\sigma= \sigma_E$ in noise model (A) and $\sigma= \sigma_{pop}$ in noise model (B).
Since the drive $\mu_E=N_E \omega_E \nu_{exc} \tau_m > 0$ corresponds to the background
activity of an excitatory population, the mean drive is always positive.
For $N_E=100$, $\omega_E=1mV$, $\nu_{exc}=1Hz$ and $\tau_m=20ms$ we get a mean depolarisation
of $\mu_E=2mV$ due to the noise input.

\subsection{Network structure}
Both excitatory and inhibitory neurons are modelled with a
membrane time constant $\tau_m$ of 20ms. The neurons are weakly
(connection probability = 0.2) and randomly connected through
simple static synapses. We carefully chose the synaptic
strengths, $\omega_I=5\omega_E$ so that for $\nu_{exc}=2-5 Hz$
and $N_E = 100$, the network is in a regime of asynchronous and
irregular firing (see figure \ref{spikes}B centre), based on the
phase diagram described in \cite{Brunel00}. In such a regime,
there is a stationary overall activity and highly irregular spike
firing of individual cells. Specifically, we take
$\tau_{rp}=2ms$, $D=1ms$, $\theta=5mV$, $u_{reset}=0mV$,
$\omega_E=1mV$ and $\omega_I=5mV$ (see table 1) , but other combinations of
parameters give qualitatively similar results. An extensive
scanning of the parameter space has been carried out\footnote{For
example, the precise values of the synaptic strengths affect the
exact location of the optimal performance when larger than 0.1mV
but not the overall trend of the results.}. This robustness
originates from the fact that the phase diagram of the type of
networks we study in this article, can be qualitatively drawn as
a function of two parameters only; the external drive (mean and variance) and the
effective ratio between excitation and inhibition. The presence
of an asynchronous phase of neural activity (and therefore the
essence of our results) is guaranteed for a broad range of the
neuron's parameters \cite{Brunel00}.

With a stronger input, the system reaches a phase of synchronous
firing with fast oscillations \cite{Brunel00}. For weak 
drive, activity will tend to zero (quiescent state)
since the network itself is not capable of sustaining intrinsic
activity.

As we will see later, the absence of synchrony at the working
point of the network is important in order to let information
about a past stimulation flow for a long time in the network.

\subsection{Evaluating information buffer properties}

We perform simulations of the network with some time-dependent
inputs $I_i^{ext}(t)$. The question we want to ask is: how much
information is left about the external inputs after some time $T$
? Given some continuous input $I_i^{ext}(s)$ we expect that the
instantaneous state of the network (given by its membrane
potentials $\{u_j(t)| 1 \leq j \leq n \}$) holds information
about $I_i^{ext}(s)$ for $s<t$. If a reconstruction of the input
signal $I_i^{ext}(t-T)$ can be achieved by looking at the membrane
potentials at time $t$, we say that the network can "buffer"
information for a time $T$.

\begin{figure}[!hbt]
\hbox{\large{A} \hspace{50mm} \large{B}}
\hbox{\includegraphics*[angle=0,scale=0.29]{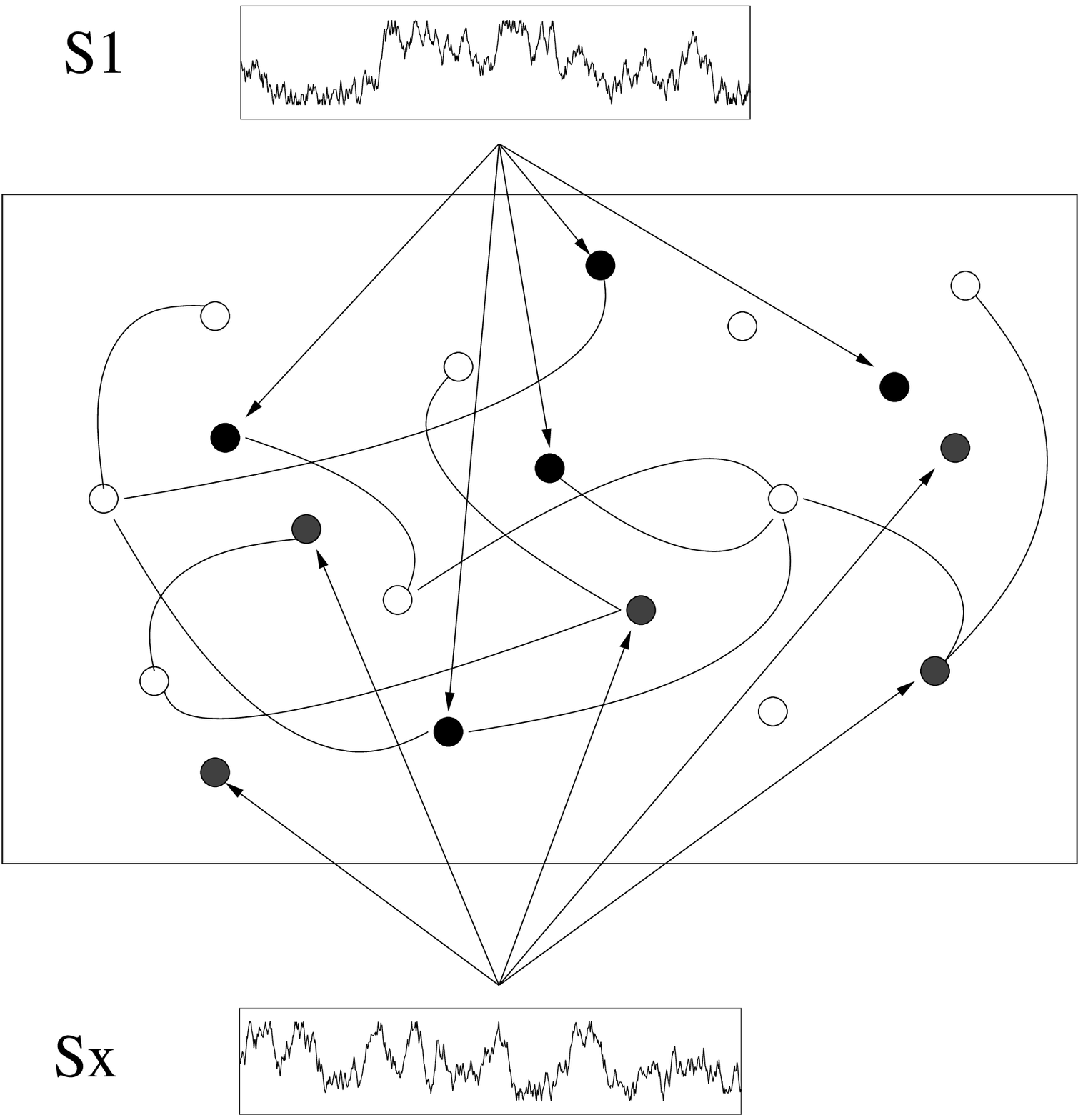}
\hspace{5mm} \includegraphics*[angle=0,scale=0.34]{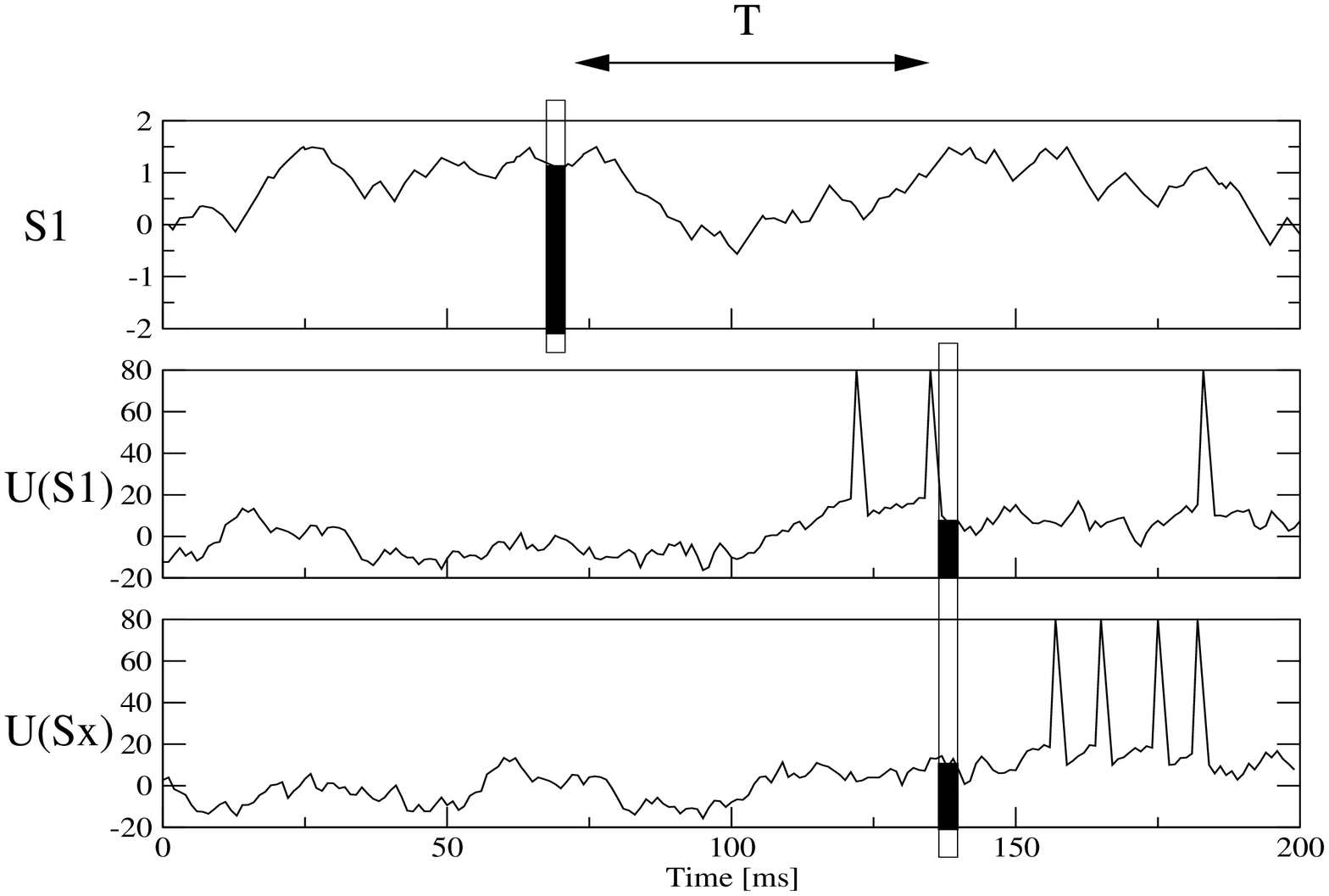}}
\caption{A: The different signals S1,..,Sx are introduced to
randomly chosen interconnected neurons. Every neuron receives
exactly one input from the exterior. Only two input signals S1
and Sx are shown in the figure. B: Based on the momentary state of
all membrane potentials (those receiving directly the signal S1
\emph{and} those receiving any other signal Sx), a readout
structure is trained to guess the amplitude of its corresponding
input a time T before (referred as delay thereafter).}
\label{fig1}
\end{figure}

We assess the information buffering capacity of the network with
a procedure analogous to \cite{MaassETAL:01a} and \cite{Jaeger02}.
We inject simultaneously $N_s$ independent input signals to $N_s$
disjoint groups of randomly chosen neurons (see figure
\ref{fig1}A), every neuron receiving exactly one input; e.g. for
$N_s=4$ input signals we have 4 groups of 50 neurons in our
network of 200 neurons. $N_s$ readout structures have access to the membrane potentials of all neurons in
the network at time $t$ and use this information to estimate the
input signal that was present at time $t-T$ (see figure
\ref{fig1}B).

We consider two different methods:
\begin{itemize}
\item{Linear read-out.}

The outputs of the readout structures are simple linear
combinations of all the membrane potentials i.e.

$$Output(t)=\sum_k \alpha_k u_k(t)+\alpha_0 = \vec{\alpha}^T \cdot\vec{u}+\alpha_0$$
where $\vec{u}= (u_1,u_2,...,u_n)^T$ is the vector of membrane potentials and $\alpha_0$ a bias term.

\item{Non-linear kernel-based read-out.}

In order to know whether more complex read-out structures can
extract significantly more information than the linear read-out,
we used Support Vector Machines (SVM's) \cite{Christianini00} , a
powerful method from the field of statistical learning \cite{Vapnik99}. 
It belongs to the family of kernel-based
methods \cite{Scholkopf02}. In those techniques, the input is sent to a
high-dimensional space in which linear fits are expected to work,
i.e.

$$Output(t)=\sum_l \alpha_l \mathcal{K}(\vec{u}(t),\vec{u}(t_l))+\alpha_0$$

The sum runs over several samples of instantaneous network states $\vec{u}(t_l)$.
The set of relevant samples is selected by the SVM algorithm.
Different kernels $\mathcal{K}(\cdot)$ can be used in order to implicitly
build a multi-dimensional space. In section
\ref{svm} we will compare the linear read-out to a Support Vector
Machine with a Gaussian kernel. We used SVM-torch \cite{Collobert01} for the SVM method.

\end{itemize}

Only the parameters (the weights $\alpha_k$) of the readout structures
are tunable, while the network of IF neurons itself remains
fixed. Since we want to assess how much we can learn about the
input at time $t-T$ if we read out the membrane potentials at
time $t$, we minimise the error :

$$E(\vec{\alpha})=\frac{1}{2A}\sum_t[Output(t)-Input(t-T)]^{2}$$

with $A=\sum_t (Input(t)-\overline{I})^2$ where $\overline{I}$ denotes 
the mean drive. Minimisation is achieved 
by an optimisation procedure (optimal regression for the linear
read-out) on the training set. Later in the results section, we
will refer to this error as the \emph{signal reconstruction
error}. The $Input(t-T)$ plays the role of a target value for
optimisation. For the linear readout there is a unique optimal
set of weights $\{\alpha_k\}$ for every delay $T$\footnote{The application of the projection theorem \cite{Luenberger97} 
say that it is true if the number $n$ of 
free parameters (the number of neurons, $n=200$ in our case) 
is smaller than the number of independent examples. 
We took 50'000 examples, but these are slightly correlated in time. 
A realistic measure is to divide the number of
examples by the effective autocorrelation of the input (3.3ms up to 80ms, see section \ref{inputprop}).
Even in the worse case, a lower bound for the number of linearly independent examples is $\frac{50'000}{80}=625>n=200$.
The lengths of the simulations guarantees that there is a unique optimal set of weights.
}. After a
training period, the weights of the readout structures are frozen
and $N_s$ new input signals are introduced in the network.
Outputs of the readout structures are then compared with their
corresponding targets. For all the simulations, the trajectories
used for training correspond to 50'000 time steps of simulation.
Performance is measured on an independent test set of 50'000
time steps. For the sake of clarity, all figures in the result
section are obtained for a single input $N_s=1$. In the
discussion section, we vary the number $N_s$ of inputs.

\subsection{Autocorrelated inputs}\label{inputprop}
As mentioned before, all neurons in the network receive one of
$N_s$ input signals $I_i^{ext}(t)$. We build an input such that its
autocorrelation profile depends on a single parameter and
decays to zero beyond a given value. In order to do so,
the total simulation time is broken into segments of duration $T_{max}$.
During each segment of length $T_{max}$ the input is kept constant.
At the transition to the next segment, a new value of $I_i^{ext}(t)$ is drawn
from a uniform distribution centred at zero, between
-20 pA and +20 pA (-7.5 pA and +7.5 pA as an additional example shown in figure \ref{poisson}B). 
This procedure results in a triangular
autocorrelation profile $A(s)$ that is zero for $t>T_{max}$, i.e.

\begin{equation}
  A(s)= \begin{cases}
    A_0(T_{max}-|s|)&   \text{for} |s|<T_{max}, \\   \notag
    0&  \text{for} |s|\geq T_{max}. \notag
  \end{cases}
\end{equation}

Thus the signal at time $t$ provides no information about the signal at $t-T$ for $T>T_{max}$.
With the triangular autocorrelation, we can compute the
effective autocorrelation $\tau_{in}$ (that can be seen as the time constant) of such an input:
\begin{equation}
    \tau_{in}=\frac{\int_0^{\infty}|s|A(s)ds}{\int_0^{\infty}A(s)ds} \notag
\end{equation}
which yields $\tau_{in}=\frac{T_{max}}{3}$.
In our simulations we explored values of $\tau_{in}$ in the range
of $3.3 < \tau_{in} < 80 ms$.

\subsection{Spectral analysis}\label{spectrum}

For the interpretation of some of our results, it is useful to consider the
power spectrum of the network activity. The spectrum is obtained
by a fast Fourier transform on the temporal evolution of the
spiking activity during $2^{16}$ time steps. The spectrum is then
averaged over $1000$ frequency steps in order to smooth the
curves. Such spectral curves give us the power at different
frequencies. A single peak at a given frequency would indicate
for example oscillations in the network at that precise
frequency. As an artifact of the averaging procedure sharp
resonance peaks would be slightly broadened. We have checked that for the simulations
shown in this paper, averaging does not cause artificial broadening.

\section{Results}
We stimulate a network of sparsely connected excitatory and inhibitory integrate-and-fire neurons
by a continuous time-dependent input current $I^{ext}(t)$.
We will evaluate for how long information about an input
signal is kept in the network.
To this end, we try to reconstruct the input signal at time $t-T$
by looking at the instantaneous membrane potentials of all neurons at time $t$.
If the reconstruction is possible, we say that the network is capable of
"buffering" information for a time $T$. The "buffering" performance as a function of $T$
is the central quantity of interest. Information buffering is possible if
the error of signal reconstruction is small.

In the first subsection we investigate the role of noise on the
performance measured in terms of the signal reconstruction error.
We show that a moderate amount of noise systematically improves
the performance. The role of asynchronous firing
is emphasised.

In the second subsection we focus on the
temporal characteristics of the input signal. We derive an empirical
relation between the temporal memory of the network and the
effective autocorrelation of the input.

Finally we show that the rich dynamics of the network in its
asynchronous state provides a good representation of its past
stimulation. Therefore simple linear readouts are powerful enough
to extract information and perform nearly as well as advanced
kernel-based methods \cite{Scholkopf02,Christianini00}.

\subsection{On the importance of noise}\label{noise_section}
\subsubsection{Stochastic arrival of excitatory background spikes}\label{section_poisson}

Networks of integrate-and-fire neurons are known to have complex
dynamics \cite{Brunel00,Brunel99}. In particular, our network
of 200 excitatory and inhibitory neurons
switches from highly synchronised neuronal activity to a very
irregular firing regime, depending on the external input; figure \ref{spikes}A and B, see also
\cite{Brunel00}. In the asynchronous phase, individual cells have
a coefficient of variation $C_V$\footnote{This is a measure of the width of the distribution of spike intervals, 
defined as the ratio of the variance and the mean squared. 
A Poisson distribution has a value of $C_V=1$. A value of $C_V>1$, 
implies that a given distribution is broader than a Poisson distribution with the same mean. 
If $C_V<1$, then the spike train is more regular than that generated by a Poisson neuron of the same rate (see for example
\cite{Gerstner02}).} 
close to or larger than 1, reflecting
this stochasticity (for the analytical computation of $C_V$, see \cite{Brunel00}). 
The asynchronous firing phase can be reached
with an external noise term corresponding to stochastic spike
arrival at excitatory synapses (noise model A, see methods). The
phase diagram of such a network is fully described in
\cite{Brunel00}.

\begin{figure}[!hbt]
\hbox{\large{A} \includegraphics*[angle=-90, scale=0.23]{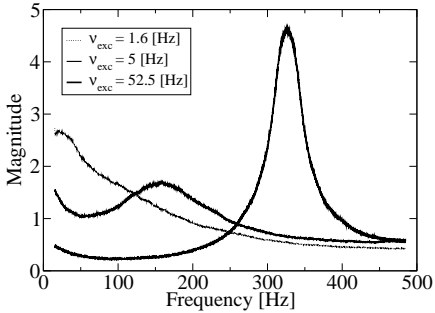}
\large{B} \includegraphics*[angle=-90, scale=0.23]{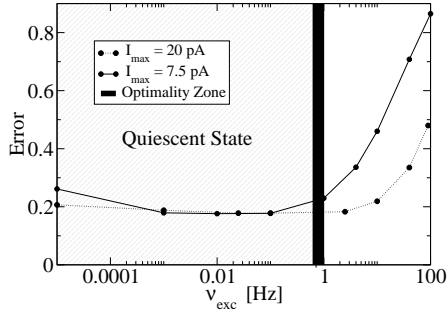}}
\caption{A: Power spectra of the population firing activity.  B: Error
as a function of Poisson noise. Notice the non-monotonic dependence on
the amount of noise.  As explained in \cite{Brunel00}, the
asynchronous firing phase can be found for an intermediate drive
(rather flat power spectrum). In a lower noise regime the neurons
stand farther from threshold, the network remains silent in absence of
a signal. Hence a spike-based readout would only become effective for
an external drive larger than $\nu_{exc} \simeq 0.8$Hz. The zone of
optimality is then located at the upper limit of the quiescent regime.
In the high excitation limit, the mean drive dominates the effect of
the "pure noise" component and oscillations appear.}
\label{poisson}
\end{figure}

\begin{figure}[!hbt]
\hbox{\large{A} \includegraphics*[angle=-90, scale=0.16]{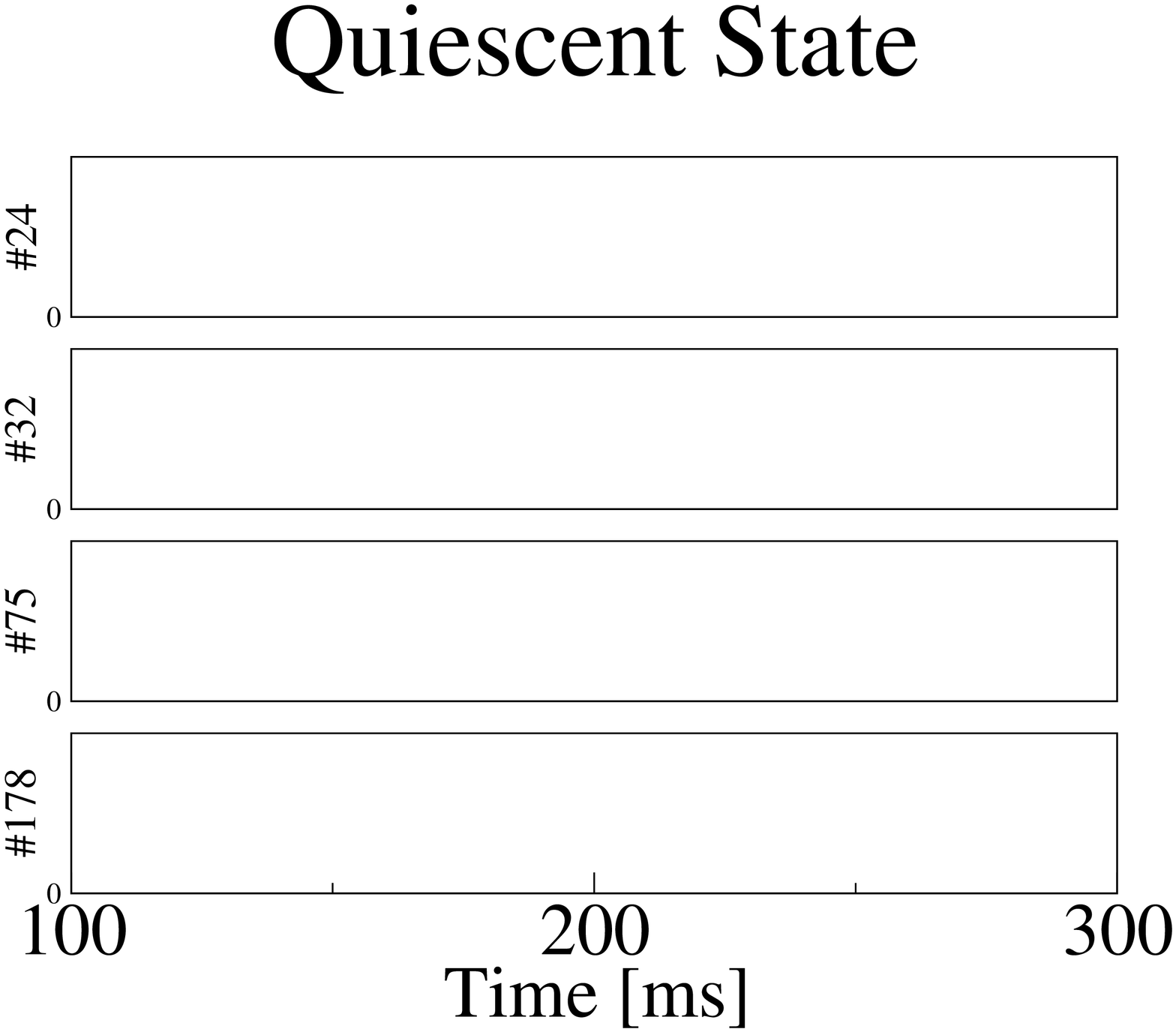}
\includegraphics*[angle=-90, scale=0.16]{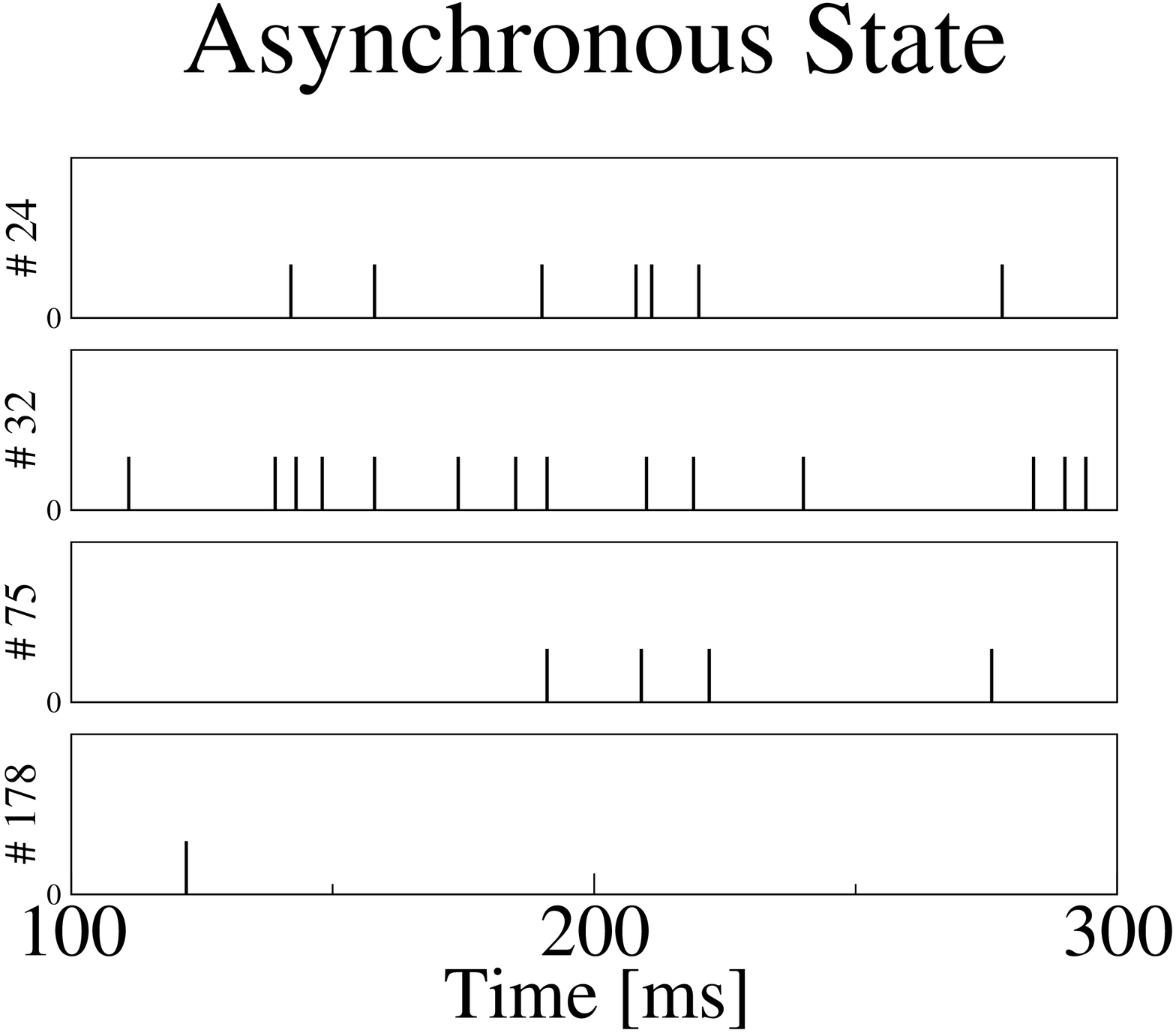}
\includegraphics*[angle=-90, scale=0.16]{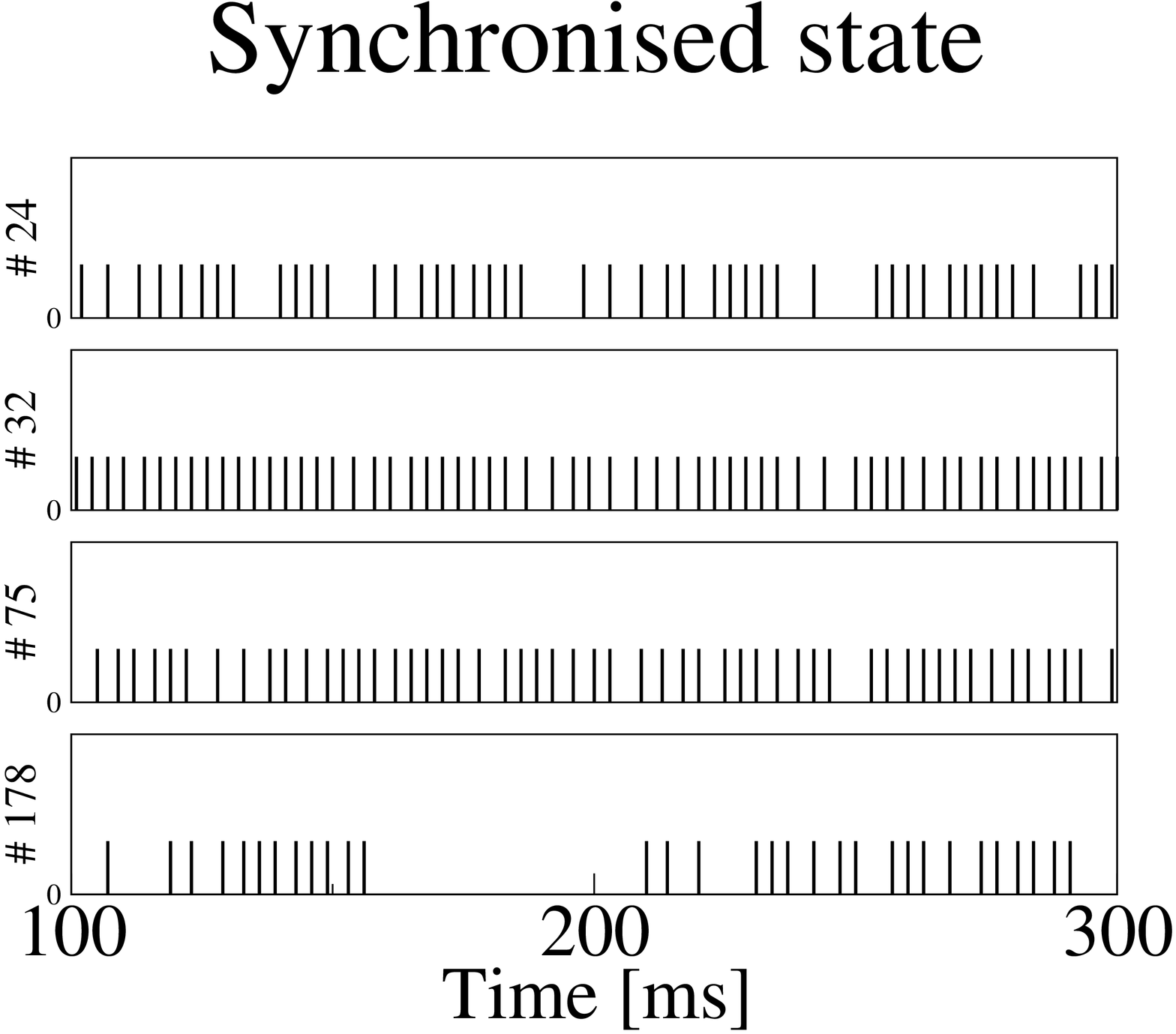}}
\hbox{\large{B} \includegraphics*[angle=-90, scale=0.16]{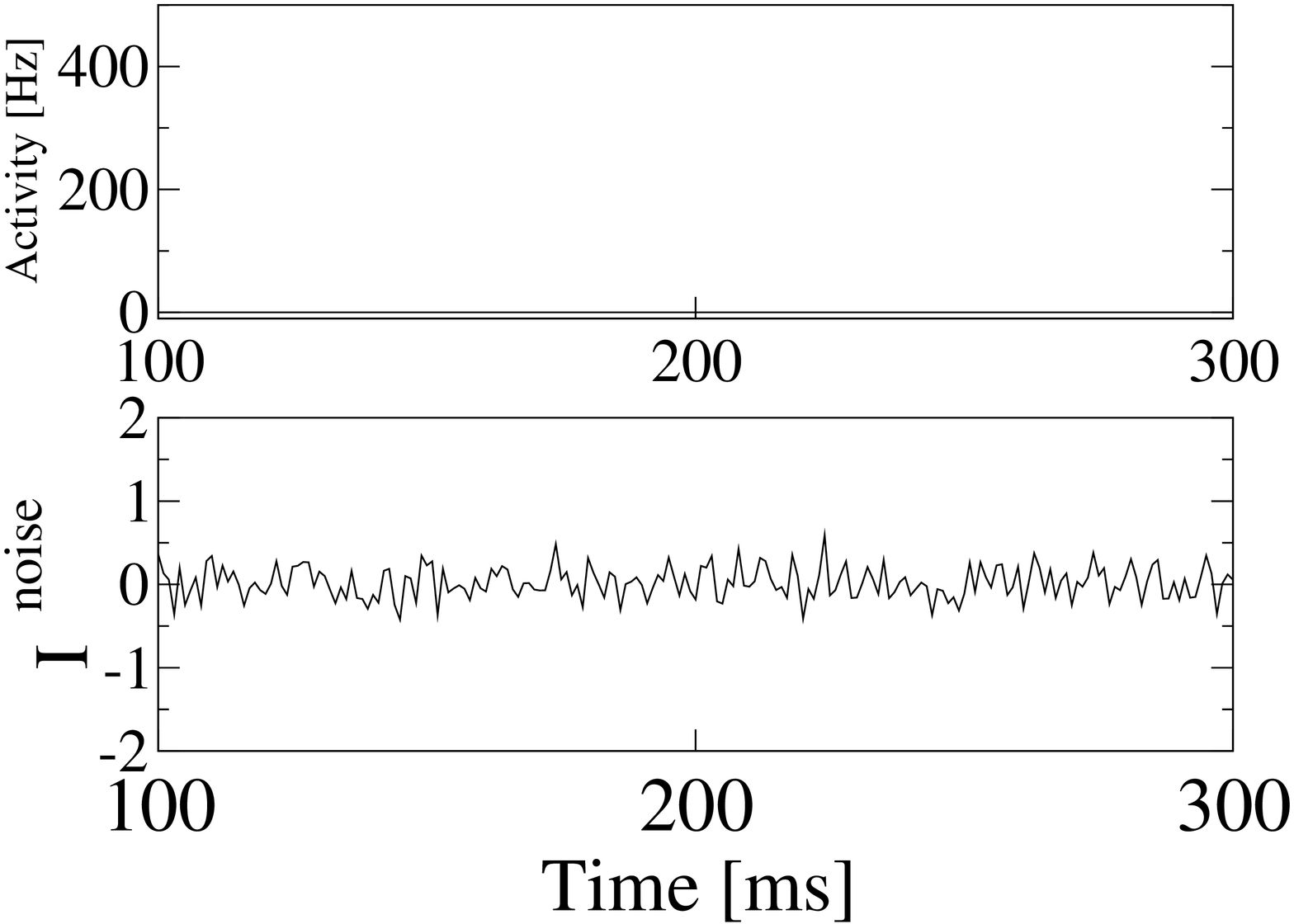}
\includegraphics*[angle=-90, scale=0.16]{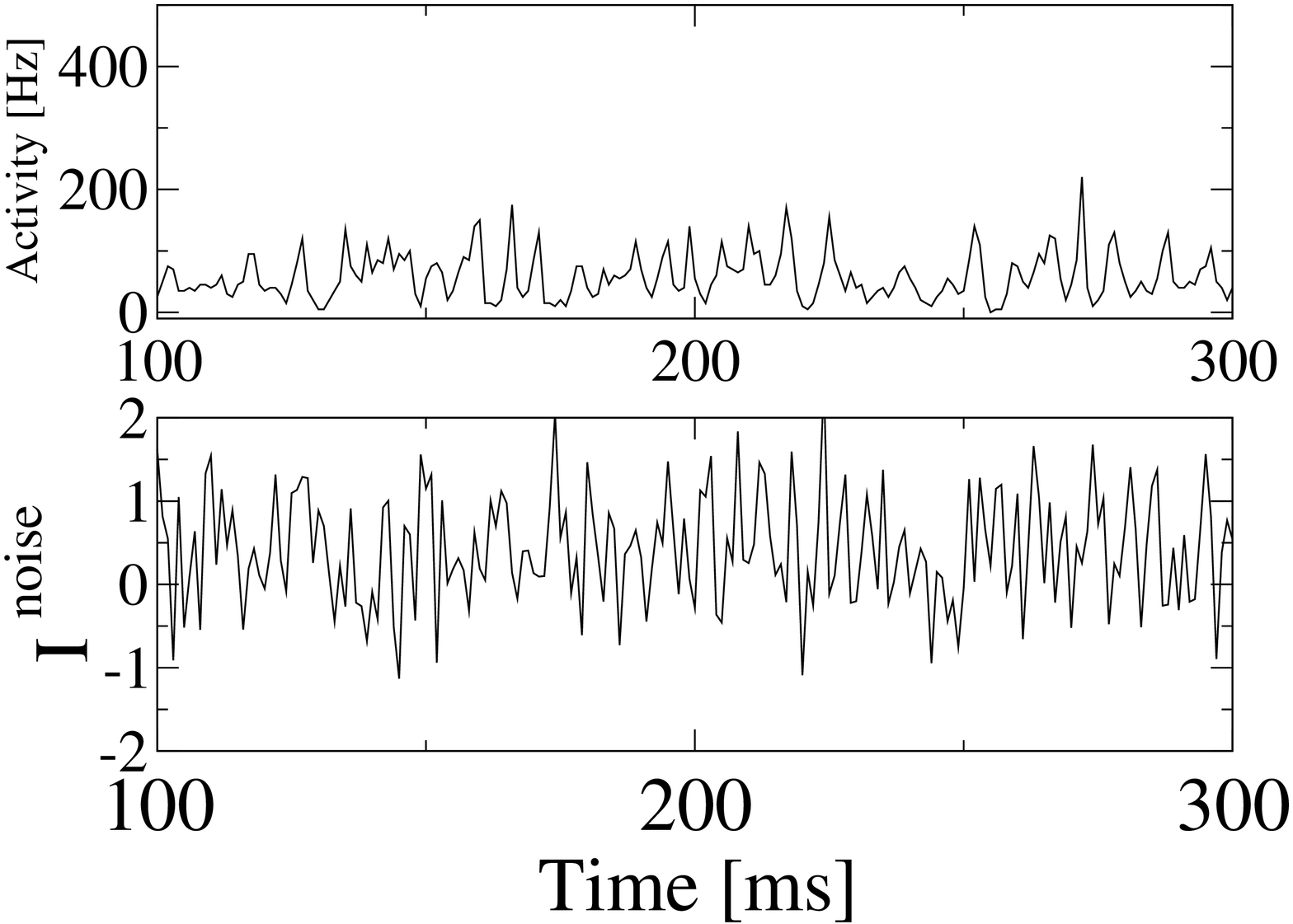}
\includegraphics*[angle=-90, scale=0.16]{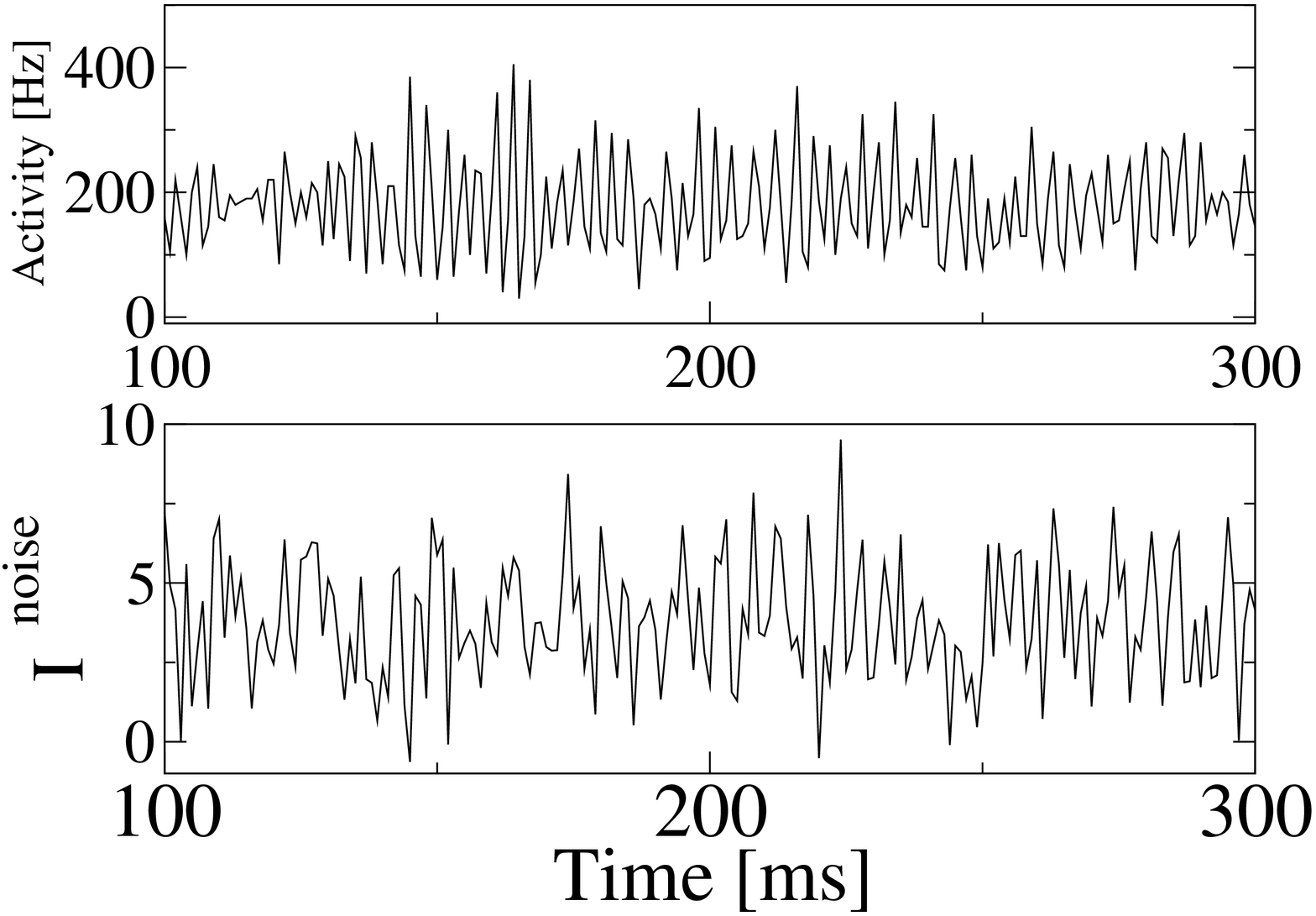}}
\hbox{\large{C} \includegraphics*[angle=-90, scale=0.16]{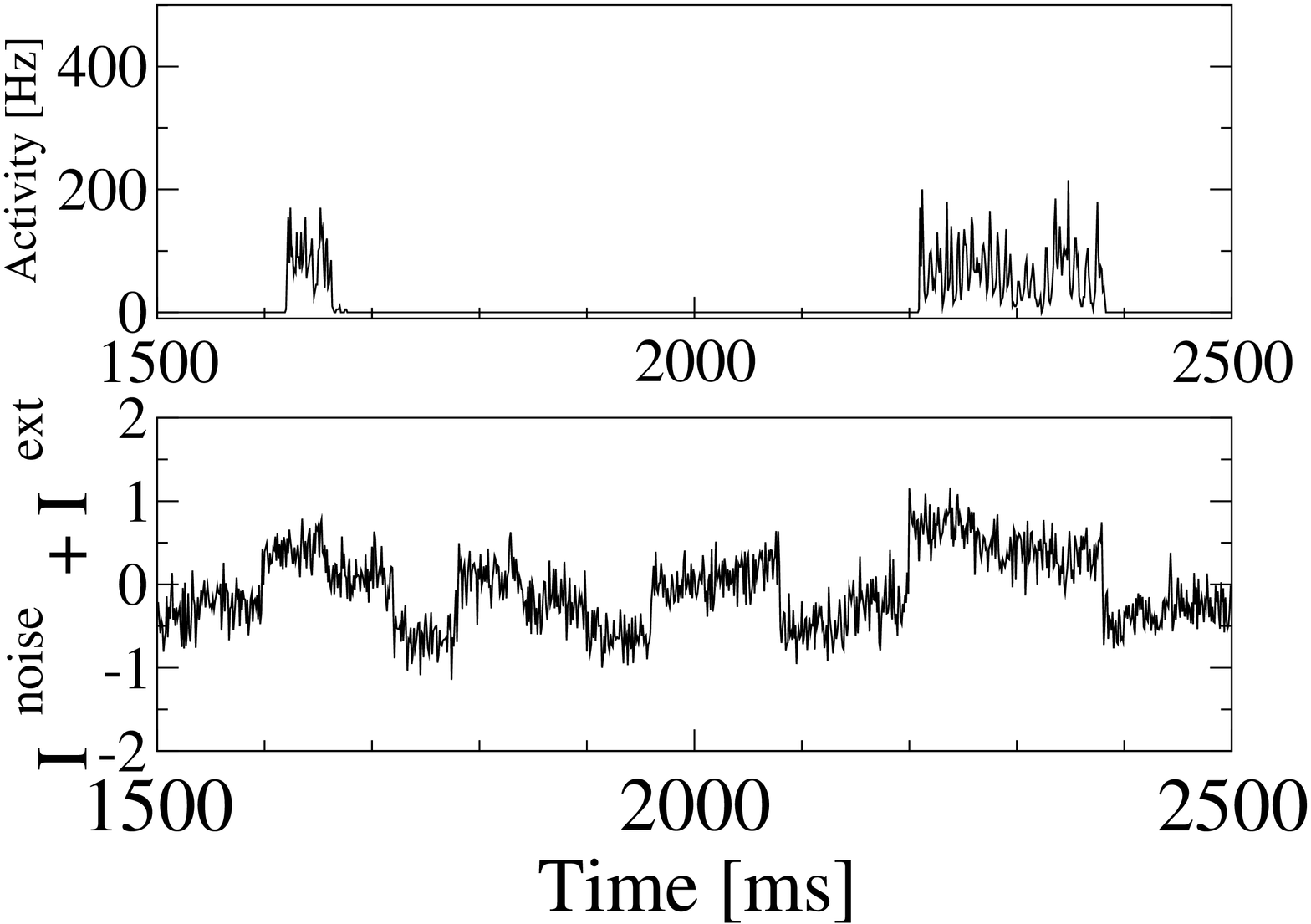}
\includegraphics*[angle=-90, scale=0.16]{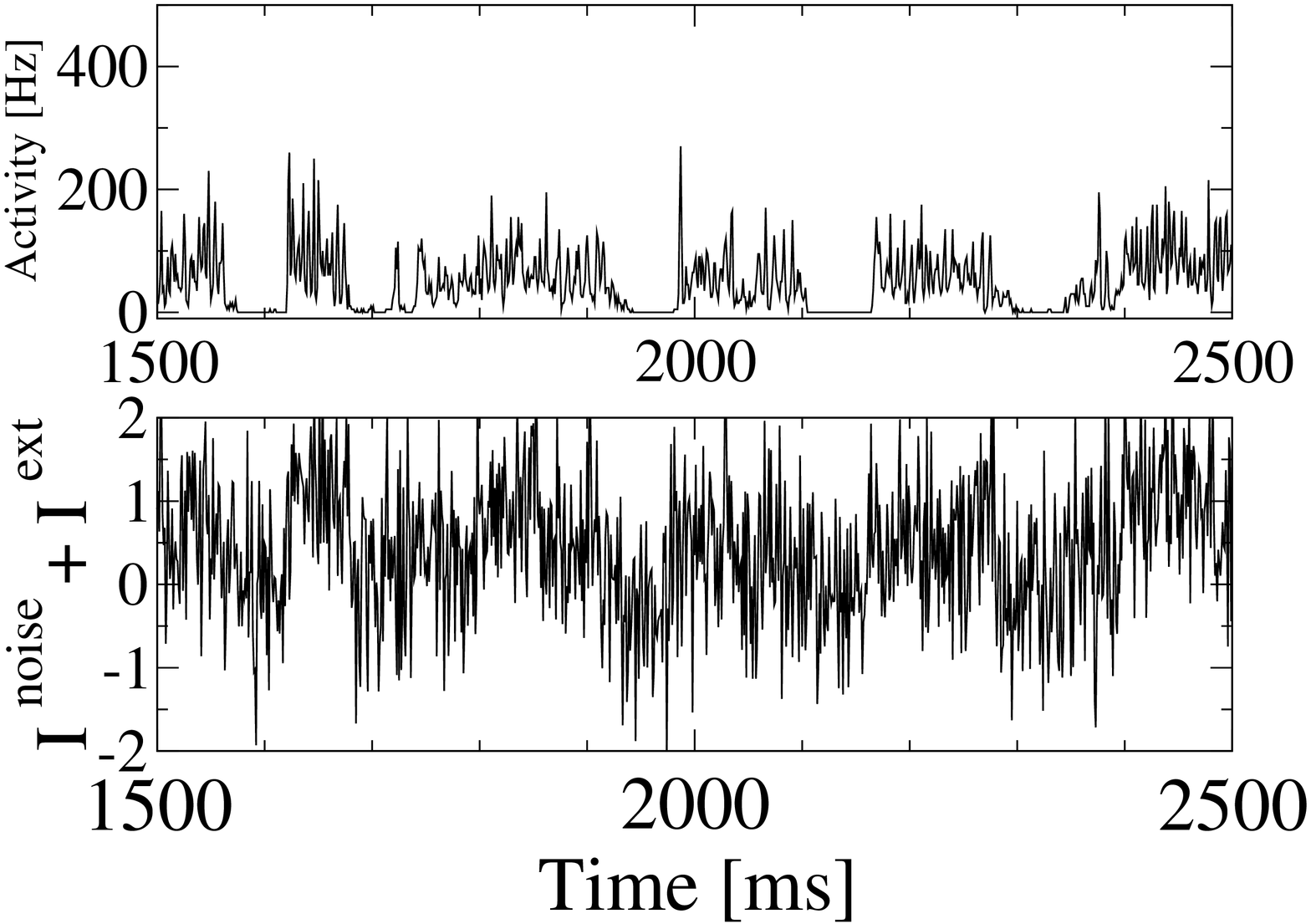}
\includegraphics*[angle=-90, scale=0.16]{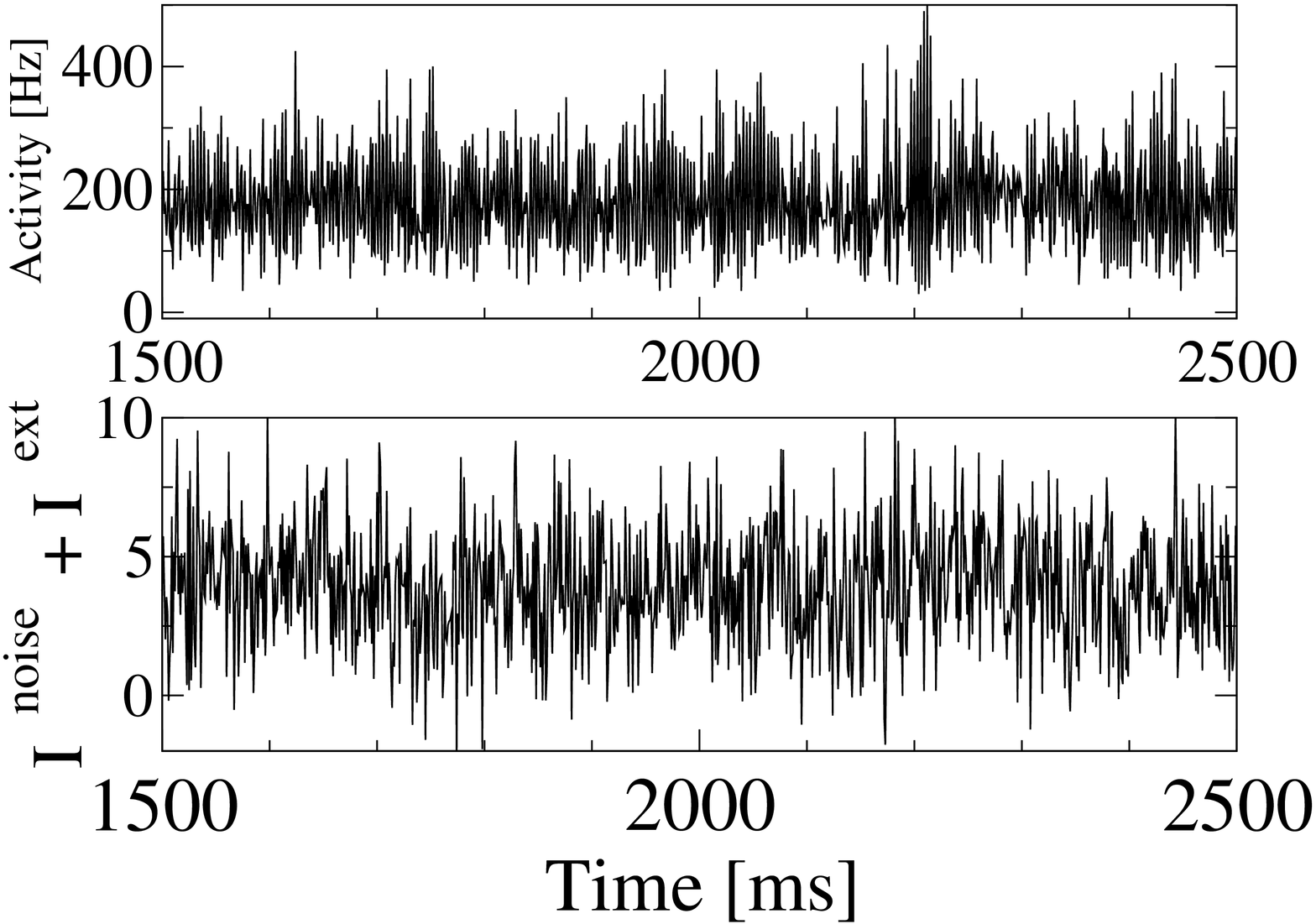}}
\caption{A : Representative spike trains in absence of the time
dependent external signal $I^{ext}(t)$, for the quiescent state
(left), for the asynchronous state (centre) and for the
synchronised state (right). B: Network activity and a
representative noise signal for the quiescent state (left), for
the asynchronous state (centre) and for the synchronised state
(right). Note the different scale for the noise in the
synchronised regime. C: The external input $I^{ext}(t)$ has been
switched on at time 1000ms (not shown on the graph). We show the
network activity and the sum of the external input signal
$I^{ext}(t)$ and a representative noise signal for the quiescent
state (left), for the asynchronous state (centre) and for the
synchronised state (right).} \label{spikes}
\end{figure}

In order to characterise the network activity with this first
model of noise (equation (\ref{baleq}) see methods), we perform a
preliminary series of simulations for different stochastic spike
arrival rates $\nu_{exc}$ at excitatory synapses in absence of a
deterministic external stimulus. The power spectrum of the
activity illustrates the regime of firing for the different
drives, in particular the regimes of fast oscillations 
(above $\nu_{exc} \cong 10 Hz$, e.g for $\nu_{exc}=52.5 Hz$, see
figure \ref{poisson}A) and irregular asynchronous firing
($\nu_{exc}\cong 1-5 Hz$). In the asynchronous firing phase, the
power spectrum has no significant resonance peak (figure
\ref{poisson}A, $\nu_{exc}=1.6 Hz$), the individual spike trains
are irregular (figure \ref{spikes}A centre) and the overall
activity has no oscillations (figure \ref{spikes}B centre) in
absence of input signal.

In a second series of simulation, we apply a time-dependent stimulus current $I_i^{ext}$. Neuronal activity
reflects the temporal structure of the input (see figure \ref{spikes}C). In order to assess
whether we can estimate the input at time $t-T$ from the set of
membrane potentials at time $t$, we measure the signal
reconstruction error (see methods).

In figure \ref{poisson}B, short-term memory buffer performance is
measured in terms of the reconstruction error on the test set for
a delay $T=10ms$ by varying the rate $\nu_{exc}$ at $N_E=100$ excitatory synapses. Information
buffer properties are good (i.e minimal reconstruction errors)
up to $\nu_{exc}\cong 2Hz$. The zone of optimality extends approximately up to the
transition zone between quiescent and asynchronous irregular
firing phase. In the quiescent regime (up to $\nu_{exc}\cong 1Hz$),
there is no activity in absence of the input signal $I^{ext}$ as
seen in the left part of figure \ref{spikes}B\footnote{A more
realistic readout, that would have access to the spikes of the
individual neurons solely, would therefore only work when the
network is not silent. Thus the zone of optimality would be
dramatically narrowed.}. However the
neurons are close to threshold and the input signal, that is
switched on at time 1000ms, make them fire when the input becomes
sufficiently large, figure \ref{spikes}C left. For a weaker
drive, neurons are farther from threshold and the input signal
has to be stronger in order to elicit a response.

As the frequency $\nu_{exc}$ is increased beyond $1Hz$, the network reaches a
phase of asynchronous firing. The population activity is now
modulated by the input signal (figure \ref{spikes}C centre).

Increasing the excitatory background input does not only increase
the noise, but also the mean drive. A resonance peak is already
present in the high frequency domain for a moderate drive (figure
\ref{poisson}A, $\nu_{exc}=5 Hz$). The increased effective drive
generates network activity at a high frequency (see the
large peak at about 300Hz in figure
\ref{poisson}A, $\nu_{exc}=52.5 Hz$). In such a regime, the
individual cells fire at a large rate, figure \ref{spikes}A
right, and the overall activity has fast oscillations in absence
of input signal, figure \ref{spikes}B right. The excitation by
the time-dependent external signal $I^{ext}(t)$ is buried in the noise of the stochastic excitatory background, and the
overall activity is barely modulated by the signal, figure
\ref{spikes}C right. The performance hence decreases dramatically
with an increase in the excitatory firing rate $\nu_{exc}$.

We can thus deduce from the error plot in figure \ref{poisson}B, that the network operates
at best close to the stochastic dynamics, rather than in a network dominated by oscillations.

Signature of chaos in neural systems is not a new concept.
It was already shown in the past that networks of model neurons can give rise to a chaotic 
firing regime \cite{Hansel96,Sompolinsky88,Vreeswijk96}.
Experimental evidence indicates that cells exhibit \emph{in vivo} very irregular spike
trains, with a coefficient of variation CV close to 1 (see e.g \cite{Softky93}), 
that may be the signature of a rich and very irregular underlying dynamics. 
From both an experimental and a theoretical perspective, the brain is thought to 
exhibit a chaotic behaviour \cite{Faure01,Faure97,Korn03}.

In parallel, the fact that information processing by a dynamical system may gain from being close to 
a chaotic regime was postulated in the framework of cellular automata \cite{Langton90}.

But only recently studies have focused on the fact that 
the brain may actually be operating at the edge of chaos
\cite{Rowe01,Bertschinger04}. Our findings suggest that networks of
\emph{spiking} neurons can efficiently buffer information when close to 
a stochastic dynamics.

Spontaneous stochastic activity at low rates is also known to be
a prerequisite of rapid signalling \cite{Gerstner00,Brunel01}.

\subsubsection{A relation to stochastic resonance: model B}\label{stochres}

\begin{figure}[!hbt]
\hbox{\large{A} \includegraphics*[angle=-90,
scale=0.23]{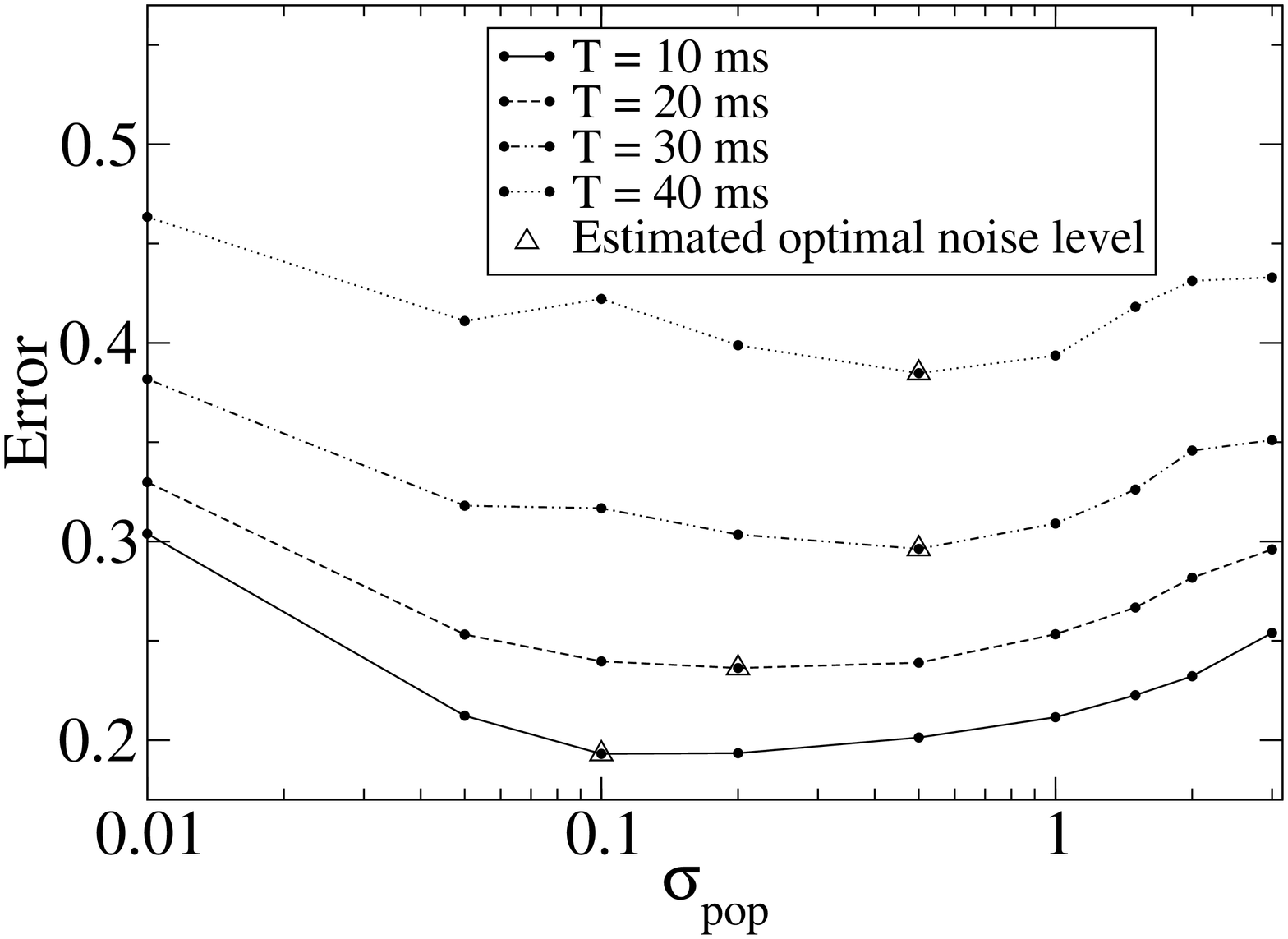} \large{B}
\includegraphics*[angle=-90, scale=0.23]{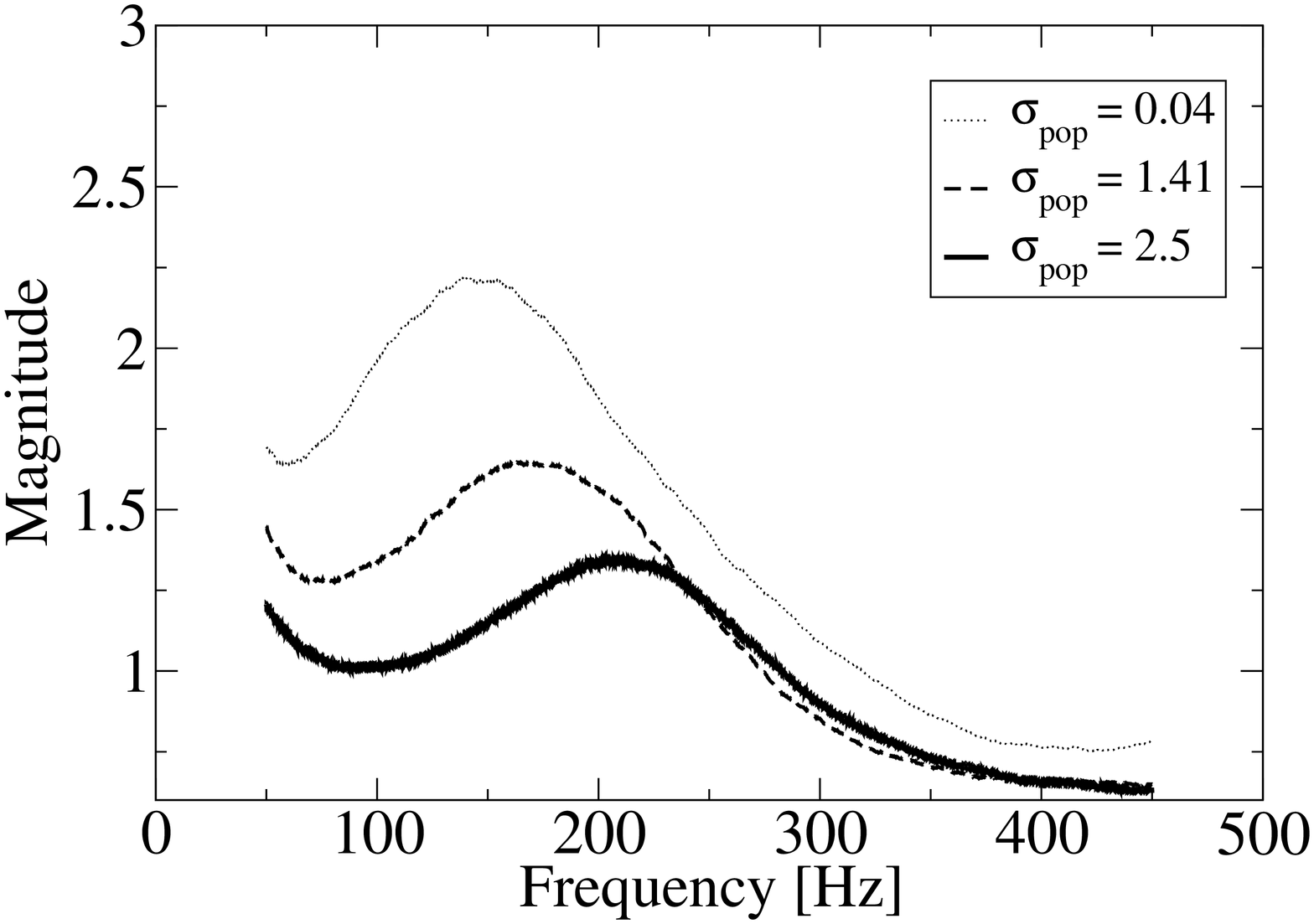}}
\caption{A: Errors as a function of the amplitude of noise in
model B. Again, we can see that a moderate amount of noise level
increases the performance of the network. In this case, this is
purely a noise effect since the mean drive is constant. Thus it
an also be interpreted as a stochastic resonance effect. Notice
there is a different optimal noise level for the different delays
$T$. A possible explanation is proposed in section \ref{stochres}.
B: Corresponding power spectra of the activity. Since the
constant drive is fixed, the noise just smoothes out the resonance
peak present at a low noise level. Even though the peak in the
power spectrum at high noise is slightly weakened, noise still
has a corrupting effect on the input. The neurons are well
desynchronised but the signal-to-noise ratio gets very low.}
\label{balanced}
\end{figure}

We can also set the network in an state of a stationary network
activity with irregular individual spike trains by driving it
with a balanced excitatory and inhibitory spike input (see
methods, noise model B). As detailed in section \ref{model2}, this
drive is equivalent to a constant input (contributed by the purely excitatory
population only) and additional white noise (the variance parts of
both the purely excitatory population \emph{and} the balanced
population). The performance is evaluated for different white
noise amplitude on the top of a constant depolarisation of
$\mu_E=0.6mV$. Errors on the test set are computed for different
delays $T$ (see figure \ref{balanced}A). As previously seen for
the noise model A, there is a non-monotonic dependence of the
performance upon the noise level.

The signal reconstruction error is relatively high when the noise
is low since the network is in a quiescent state. Addition of
noise sets up the network in a state where the neurons are weakly
coupled; each neuron act as an almost independent unit and its
coupling to the other neurons enriches the network dynamics: we
have reached a state in which the neurons are optimally
desynchronised. By adding more noise, the signal-to-noise ratio
decreases along with the ability to retrieve information. The
resonance peak present in the low noise case is smoothed out as
the noise level increases\footnote{In the case of a larger
network, the resonance peak would be washed out. Since our
simulations include a small number of neurons, finite size
effects (such as the impact of fluctuations) become strong.} (see
figure \ref{balanced}B). In contrast to section
\ref{section_poisson}, this non-monotonic dependence on the drive
is now a pure noise effect, since the mean drive is kept constant.

An additional interesting effect can be seen in figure
\ref{balanced}A. For one delay $T$ to another, the signal
reconstruction error is at its lowest for a different noise level
(referred as "optimal"). Surprisingly this optimal noise
amplitude increases with the buffering time $T$. It means that if
we want to set up the network so that information remains
available for a buffering time of 40ms, the noise has to be
stronger than if we want to extract faithfully information after
a buffering time of 10ms. A possible explanation is that in the
low noise limit, transients are reflected almost instantaneously
in the activity profile \cite{Gerstner00,Knight72b}. A low noise
level thus allows the readout to have a fine temporal
resolution.  In the high noise limit however, the network acts as
a low pass filter \cite{Gerstner00,Brunel01}. Slow transients for
high noise allow longer buffering times. Therefore the amplitude
of noise has to be higher for long buffering times than for short
ones.

Our results are related to the well-known phenomenon of stochastic
resonance. Such an effect is seen when the response of a system
to a drive depends non-monotonically on noise, with an optimum at
a moderate, non-zero, noise level. There are many
pointers in the literature to physical evidences of stochastic
resonance in physical systems
\cite{Wiesenfeld98,Collins96,Gammaitoni95,Gammaitoni98,McNamara89,Wiesenfeld94,Wiesenfeld95},
and in models of neurons
\cite{Bulsara91,Longtin93,Plesser97,Plesser00}. In living systems
stochastic resonance has been reported in crayfish mechanoreceptors
\cite{Douglass93}, the cricket cercal sensory system
\cite{Levin96}, neural slices
\cite{Gluckman96}, hippocampus \cite{Yoshida02}, and the cortex
\cite{Mori02}. More importantly the brain appears to actually
make use of stochastic resonance at a functional level
\cite{Kitajo03,Hidaka00}.

\subsection{Impact of autocorrelated inputs}\label{ac}
When the input is correlated in time, the readout can extract
information from both the information buffered in the recurrent
activity and from the input itself (filtered by the neurons).
Thus if the autocorrelation of the input is short compared to
the intrinsic temporal trace of the network, the role of the
recurrency is crucial whereas for very long autocorrelation,
the output will predominantly read "online" the input, making use
of the "self-memory"\footnote{On average the signal at time $t$
holds information about itself at time $t-T$ for $T<T_{max}$.} of
the stimulation signal. We expect then to have an increase in
performance as a function of the autocorrelation of the input.
We can rule out an alternative explanation that a long autocorrelation 
has more time to impact the dynamics of the neural network by comparing 
the cross-correlation of the reconstructed output and the input with the 
signal autocorrelation for different autocorrelation profiles 
(for an illustrated example and additional arguments, see \cite{Mayor03}). 
Thus the cross-correlation depends on the shape and on the duration $\tau_{in}$ of the signal autocorrelation.

\begin{figure}[!hbt]
\hbox{\large{A} \includegraphics*[angle=-90,
scale=0.23]{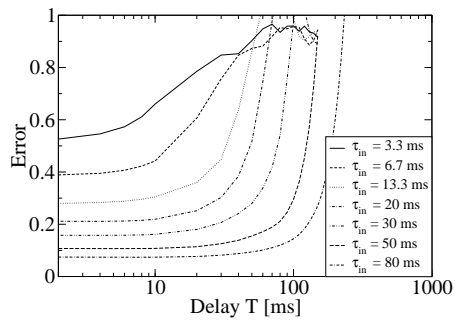} \large{B}
\includegraphics*[angle=-90, scale=0.23]{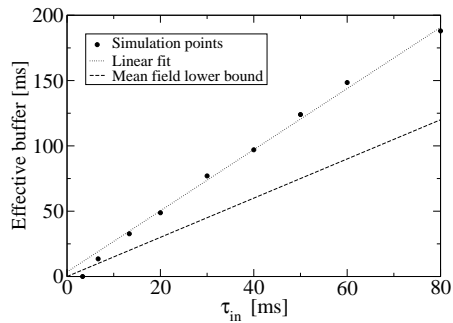}}
\caption{A: Error plots as a function of time for different input
statistics. An evident increase in performance is seen as the
autocorrelation of the input gets longer. B: Delays corresponding
to an error of 0.5 as a function of the effective autocorrelation
of the input. A linear fit shows us the clear dependence of the
persistence of information in the network on the "self-memory" of
the input.} \label{ac1}
\end{figure}

We performed numerical simulations using different effective
autocorrelation times $\tau_{in}$ for the input signals. The
results are shown in figure \ref{ac1}A. We can estimate the
effective information buffer by looking at the buffer time $T$ at
which the error rises to a value "arbitrary" chosen to be $E=0.5$ 
(the similarity of the reconstructed output to the target is still visually evident).

As we can see in figure \ref{ac1}B, the simulation points can be
fitted to first approximation by a straight line.
In the limit of long autocorrelation $\tau_{in}\gg\tau_m$ we expect from simple mean
field considerations, the buffer to have a
temporal extension longer than $\frac{3}{2}\tau_{in}$, and to approach this
line asymptotically from above\footnote{In order to derive a lower bound we could use the averaged
membrane potential for reconstruction instead all the membrane potentials
of the 200 neurons in the network.
Since the distribution of the membrane potentials is
modulated by the external input $I^{ext}$, a reconstruction based on
the average is possible.
A trivial reconstruction can then be done assuming that the
external input has not changed during the last time interval
$T$. Assuming that the network can trivially buffer a
signal that has not changed in the last time interval $T$, and that
the network cannot reconstruct at all an input that has switched in
this last time interval $T$, the effective buffer of the network
for an input that changes every $T_{max}$ is
$T<\frac{1}{2}T_{max}$. Since the effective autocorrelation of
our input is $\tau_{in}=\frac{1}{3}T_{max}$, the network has an
effective buffer of $T=\frac{3}{2}\tau_{in}$}. Indeed the
simulation results for large $\tau_{in}$ stay always above the
lower bound represented by the dashed line in figure \ref{ac1}B.

For short autocorrelation profiles (of the order of the membrane
time constant $\tau_m=20ms$ or shorter), this approximation no
longer holds, since the filtering effects due to the neuron's
membrane time constant comes into play. Rapid changes of the input on the
scale of $\tau_{in}\ll\tau_m$ will be averaged out irremediably
by the integration with the membrane time constant. Hence the
network cannot buffer efficiently (i.e. with an error smaller than
$E=0.5$) input signals that have an effective autocorrelation
$\tau_{in}$ smaller than about 5ms (see figure \ref{ac1}B).

The performance can be improved for different autocorrelation
profiles and for different network connectivities as proposed in
\cite{White04}.
They can also be increased by adding other time scales in the network,
for example at the level of the synapses \cite{MaassETAL:01a}.

Analysis of cross-correlation and autocorrelation profiles
(more precisely the analysis made in \cite{Mayor03} of the location of the peak in 
the cross-correlation profile that is shifted in proportion of the autocorrelation length)
show that the readout effectively makes an optimal trade-off
between retrieving information from the the autocorrelation of
the signal and from the buffer provided by the network
recurrency.

\subsection{Simple linear readouts are efficient}
\begin{figure}[!hbt]
  \begin{center}
  \includegraphics*[angle=-90, scale=0.4]{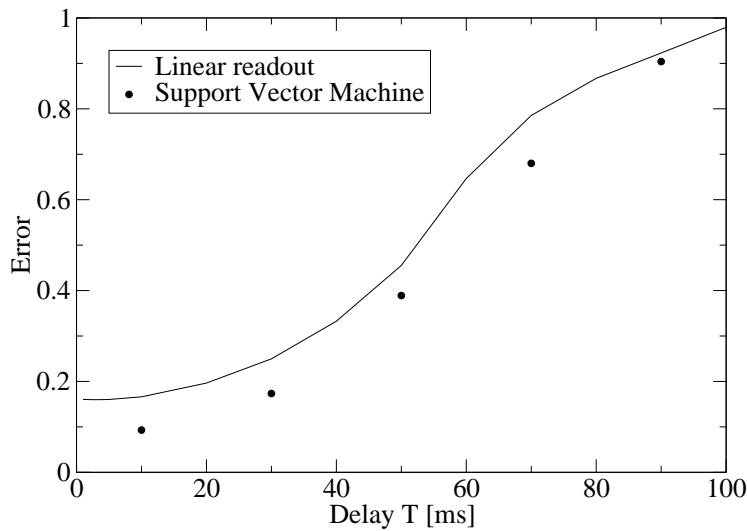}
  \end{center}
  \caption{Comparison of performances between Support Vector Machines (Gaussian kernel) and the simple linear readout on a test
  set. Optimisation has been performed on a separate training set of 50000 time steps. A simple
  linear combination of the membrane potentials is extremely effective in extracting information and is comparable to that of a
  Support Vector Machine.}
  \label{svm}
\end{figure}

As we have seen in the section \ref{noise_section}, injection of
noise in the network helps in desynchronising the neurons. The
resulting dynamics of the network is thus extremely rich. It has
been shown that in the case of formal neural networks
\cite{Bertschinger04} and analogously for cellular automaton
\cite{Langton90}, the computational abilities of complex systems
is at best when the system is close to a chaotic behaviour, as
opposed to a dynamic state with short limit cycle. Another way of
interpreting this complex behaviour is to notice that the network
provides a broad set of different filters for the input. In other
words, the one-dimensional input signal $I^{ext}(t)$ undergoes a
"dimensional blow up". This is a strong correspondence to the
working principle of some of the most efficient classification
methods, named Support Vector Machines (SVM) \cite{Christianini00}. These methods,
known under the generic name of kernel methods \cite{Scholkopf02}, send
the input to a high dimensional space. In this multidimensional
space, a simple linear separation (a hyper-plan) into two
sub-spaces is able to perform the classification. An adapted
version of those methods can handle regression problems. In
figure \ref{svm} a comparison of the performances of both the
simple linear readout and the SVM with a Gaussian kernel is
represented (see methods).
Although the kernel method achieves slightly better results than the linear combination, the gain is of the order of a few
milliseconds only.
The explanation is that an additional dimensional blow up that is at the basis of kernel methods is not necessary since a
sufficiently high dimensional representation is already given by the network itself. Hence a simple linear combination
(corresponding to the hyper-plan in the kernel methods) is able to extract most of the information needed.

\section{Discussion and conclusions}

We investigated the buffering capacity of networks of excitatory and inhibitory integrate-and-fire neurons.
The population firing activity of such a network with sparse connectivity can switch from an oscillatory firing state
to an asynchronous irregular firing regime, depending on the rate of stochastic background spike arrival.

We found that the buffering capacity of the network is optimised in the presence of a moderate, non-zero, amount of noise.
This non-trivial dependence upon noise can be related to the well-known phenomenon of stochastic resonance.
The optimal noise level, or equivalently the optimal discharge rate of an external population of neurons, corresponds
approximately to the region of transition between a quiescent state and an asynchronous irregular firing regime.
In this region, a very complex dynamics emerges from the network's activity. It provides a rich representation of the inputs.
Simple adaptable linear readouts are thus able to extract the buffered information in an efficient way.
Hence our results provide an interpretation for the role of non-oscillatory dynamics in a simplified model of cortical micro-circuits.

\begin{figure}[!hbt]
  \begin{center}
  \includegraphics*[angle=-90, scale=0.4]{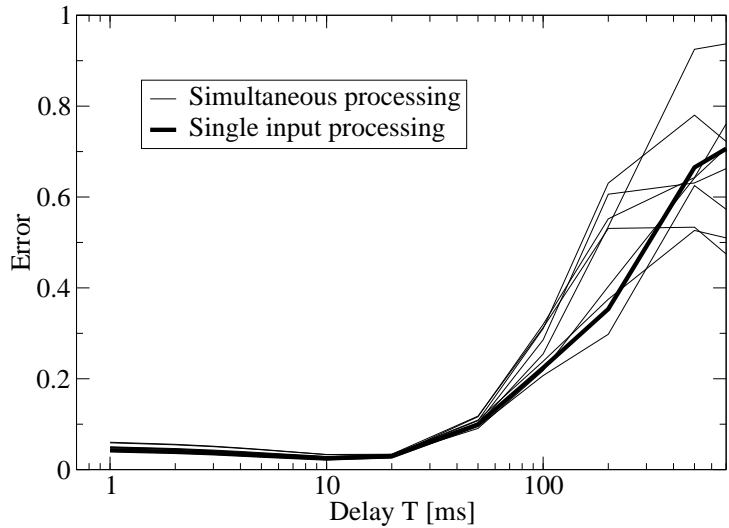}
  \end{center}
  \caption{Error plots for simultaneous buffering of eight inputs. A comparison to the
  buffering of a single input (thick line) emphasises the fact that no significant degradation
  in performance is seen for a reasonable number of inputs. Our simulations show that a
  degradation is seen only from about sixteen simultaneous inputs.}
  \label{8inputs}
\end{figure}

The buffering capacity of a such network can be generalised to the processing of a larger number of inputs.
Simultaneous buffering of eight inputs is shown in figure \ref{8inputs} in comparison to a single input.
No significant degradation is seen for a moderate number of inputs. A loss in performance can be seen only when
processing more than sixteen inputs (data not shown).
This ability to handle multiple simultaneous inputs, along with the way information can be extracted from both the
autocorrelation of the input signals and the buffer provided by the network recurrency, 
could provide an interesting way of combining
different sources of information having different time scales.
On somewhat more speculative grounds, we can think of the cortex as being in a state that is close to a chaotic behaviour, and that 
other neighbouring neural micro-circuits or even farther afferents may be
tuning the \emph{amount} of chaos (by changing their spiking rate for example) 
of a given cortical micro-circuit depending on the relevent task.
One can also imagine to build artificial networks made out of such model neurons, in order to do both information 
processing and prediction of (even chaotic) time series. The present study would suggest to set up such a device in 
a state where its dynamics can easily switch to a more complex behaviour, depending on its drive.

\end{document}